\documentclass{andp2012}
\usepackage[english]{babel}

\title{All-optical transmission modulation due to inelastic interactions of ultrashort pulses in a disordered resonant medium}

\author[D.\,V. Novitsky]{Denis~V.~Novitsky\inst{1,2,3,}\footnote{Corresponding author\quad E-mail:~\textsf{dvnovitsky@gmail.com}}}
\author[A.\,S. Shalin]{Alexander~S.~Shalin\inst{1}}
\address[1]{ITMO University, 49 Kronverksky Pr., St. Petersburg
197101, Russia}
\address[2]{B.~I.~Stepanov Institute of Physics, National
Academy of Sciences of Belarus, 68 Nezavisimosti Avenue, Minsk
220072, Belarus}
\address[3]{Saint Petersburg Electrotechnical University
``LETI'', 5 Prof. Popova Str., St. Petersburg 197376, Russia} 

\shortauthors{D.~V.~Novitsky and A.~S.~Shalin} 

\begin{abstract}
Random resonant media being one of the possible realizations of disordered metamaterials open a room of opportunities for achieving new fundamental effects and designing advanced nanophotonic devices. Strongly nonlinear optical properties of such media attract ever increasing attention nowadays from both theoretical and experimental points of view. Hereinafter we consider the case of the photonic-crystal-like structure with a randomly varying light-matter coupling provided by the random density of quantum particles. Using numerical solution of the Maxwell-Bloch equations, we study the effects of the pulse collisions in the medium. It is shown that disorder enables the qualitative changes of the system's response for copropagating pulses, whereas this is not the case for the counterpropagating ones. We propose the scheme for an all-optical transmission modulation due to the disorder-induced inelasticity of collisions of co-propagating pulses. The ability of the precise tuning the modulation via the inter-pulse distance and background refractive index adjustment is revealed. This novel approach for light control could be utilized for some highly demanded applications, such as modulation and switching of a pulsed radiation.
\end{abstract}
\shortabstract 

\begin{document}
\maketitle

\section{Introduction}

Photonic structures are usually composed in a periodical manner, e.g., multilayers or lattices of scatterers. Disordered photonics \cite{Wiersma2013} gets over this common wisdom and uses the intentionally introduced randomness as a source of novel effects and applications. The first noteworthy effect in this field of research is the Anderson localization of light \cite{Segev2013} studied also in disordered photonic crystals \cite{John1987, Schwartz2007, Vlasov1999}, photonic-crystal waveguides \cite{Garcia2017}, photonic-crystal cavities \cite{Vasco2018}, and metamaterials \cite{Gredeskul2012,Scheinfux2017}. It was recently found that disorder can induce topological state transitions \cite{Liu2017}, unexpected transmission enhancement \cite{Scheinfux2016}, and wavefront shaping \cite{Jang2018} in metamaterials. However, there are only few works employing nonlinearity in disordered photonic systems to reach such highly demanded effects as optical bistability \cite{Shadrivov2010,Yuan2016}. In this paper, we are aimed to elaborate a method of transmission modulation based on controlling ultrashort pulse propagation in disordered resonant metamaterials.

Modulation and switching of an optical response are the basic effects
needed for the controlling light signals and the construction of
tunable and bistable optical devices \cite{Gibbs,Soref2018}. A number of mechanisms and materials are proposed to realize
these effects including metamaterials and photonic crystals \cite{Shalin2015,Chebykin2015,Slobozhanyuk2015}. The first idea on the utilization of nonlinear periodic media for the light modulation had been proposed in the 1970s (see \cite{Brown1998} for a brief historical introduction). From the beginning of the 1990s, the switching and bistability
were actively studied in nonlinear Bragg reflectors \cite{Acklin1993}
and especially in photonic crystals
\cite{Scalora1994,Lidorikis2000,Soljacic2002}. Nonlinear
photonic-crystal waveguides with defects forming high-quality
microcavities \cite{Yanik2003APL} and crossings \cite{Yanik2003OL}
were also proposed for the switching realization. Generally, the effect of the
switching in photonic crystals is usually based on the band-gap
shift effect. Such all-optical modulators and switchers
utilize Kerr nonlinear response of solid materials, especially
semiconductors, such as GaAs \cite{Ermolenko2010}, zinc
chalcogenides \cite{Stankevich2005,Ermolenko2009}, InGaAsP
\cite{Nozaki2010}, and doped metal oxides \cite{Paterno2018}).
The dynamics of charge carriers in the semiconductors provide picosecond
response times of such switching systems. Recently, semiconductors
were also proposed as a key element for compact tunable metasurfaces
where the switching is the result of the nonlinearly shifting Mie-type resonances of
metaatoms \cite{Shcherbakov2017}. Note that there are plenty of other materials, which could be used as a component of photonic-crystal switching devices, such as, e.g., silica
glass \cite{Kabakova2010}, polymers \cite{Meng2012}, quantum wells
\cite{Coriasso1998}, Raman gain media \cite{Arkhipkin2014}, and transparent
conductive oxides \cite{Ferrera2018}. It is also necessary to mention the possibility to switch signals with mechanically introduced nonlinearities \cite{Shalin2014}.

Collisions of solitons and pulses in general in nonlinear media
could be also utilized for the controlling light with light including the
all-optical switching and modulation. There are many works devoted
to the interactions between pulses in a variety of nonlinear media, and the media
with Kerr (cubic) nonlinearity are perhaps the most studied ones. A
number of effects related to collisions of pulses in such media were
predicted: focusing due to the interference of spatial solitons
\cite{Cohen2002}, frequency and velocity tuning under total
reflection conditions \cite{Lobanov2010}, tunneling and trapping
regimes \cite{Sukhorukov2012}, etc. More exotic situations were also considered recently, e.g., supercontinuum pulse collisions \cite{Liu2010} and interaction between few-cycle pulses
\cite{Gao2018}.

A separate class of optical nonlinearity being of high demand for the aforementioned applications is provided by a resonant or
near-resonant light-matter interaction, when the nonlinearity is due to the
quantum transitions between the levels of active particles. One of the most intensively studied cases is a two-level resonant medium providing plenty of different nonlinear effects such as, e.g., self-induced transparency (SIT) solitons \cite{McCall,Poluektov},
unipolar pulse generation \cite{Pakhomov}, and ultrafast optical
switching due to near--dipole-dipole interaction between the particles
\cite{Crenshaw1992,Scalora1995}. As to pulse collisions, it is known that in such media,
copropagating coherent pulses eventually transform into standard $2
\pi$ solitons and interact elastically, i.e., without losing energy during the collision \cite{Novitsky2011}. Therefore, studies of the interaction between pulses in resonant media
are mostly focused on the counterpropagating scheme, in which pulses
interact inelastically \cite{Afanas'ev1990,Shaw1991,Novitsky2011}.
In particular, this fact is used to demonstrate peculiar all-optical
diode action \cite{Novitsky2012} and formation of population density
gratings \cite{Arkhipov2016,Arkhipov2017SR,Arkhipov2017OS}.

In our recent paper \cite{Novitsky2018}, we have introduced the concept of a disordered resonant medium being a kind of strongly nonlinear metamaterials. We have studied the localization of individual coherent pulses in that artificial structure. In this paper, we propose a novel mechanism for all-optical
transmission modulation based on an intense disorder-induced interaction of
copropagating coherent pulses in the disordered resonant medium. We show that in
contrast to the uniform resonant medium (no disorder)
providing elastic interaction of copropagating pulses \cite{Novitsky2011}, the
randomness of the quantum particles density (or light-matter
coupling) enables an effective inelastic interaction between the pulses. As a result, a portion of light energy gets absorbed by the
medium. This light-trapping effect strongly depends on the disorder degree
being crucial for this peculiar effect. Launching
either one or two pulses into the disordered resonant medium, one will
obtain sufficiently different levels of transmission providing an all-optical way 
for a light modulation and switching. The switching time can be estimated as the time needed for the pulses to propagate through the medium ($\sim 10$ ps for the parameters used in our calculations), whereas the return to the initial state is governed by the relaxation times of the medium.

%The paper is structured as follows. First, we introduce our model
%and discuss its main equations in Section \ref{eqs}. The main
%results are given in Section \ref{vacuum} where the possibility of
%transmission modulation due to interaction of copropagating pulses
%is demonstrated in the framework of our model. Section \ref{backgr}
%contains the study of influence of a number of parameters on the
%effect of modulation with the goal to optimize it. A short note on
%counterpropagating pulses is given in Section \ref{counter}. A brief
%conclusion summarizes the article.

\section{Governing equations}\label{eqs}

\begin{figure}[b]
\includegraphics[scale=0.5, clip=]{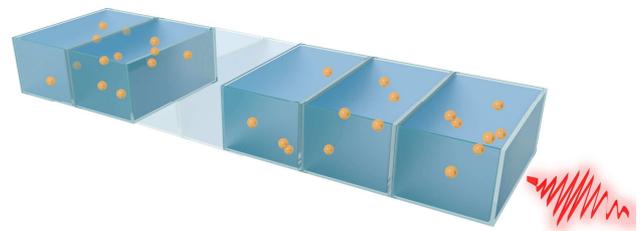}
\caption{\label{fig1} Schematic of the system
consisting of the background dielectric doped with quantum
(two-level) particles distributed with randomly changing density along the
direction of the light pulse propagation. The total length of the system
is $L$, whereas the length of the constant density layer is $\delta L$.}
\end{figure}

The system (Fig. \ref{fig1}) consists of a background dielectric doped with quantum
(two-level) particles, which initially are in a ground state. Light
propagation in this medium is described by the well-known system of
semiclassical Maxwell-Bloch equations including differential
equations for the dimensionless electric-field amplitude
$\Omega=(\mu/\hbar \omega) E$ (normalized Rabi frequency), complex
amplitude of the microscopic polarization $\rho$, and difference between
populations of ground and excited states $w$ \cite{Allen} (we use the notation from Ref. \cite{Novitsky2011}):
\begin{eqnarray}
\frac{d\rho}{d\tau}&=& i l \Omega w + i \rho \delta - \gamma_2 \rho, \label{dPdtau} \\
\frac{dw}{d\tau}&=&2 i (l^* \Omega^* \rho - \rho^* l \Omega) -
\gamma_1 (w-1),
\label{dNdtau} \\
\frac{\partial^2 \Omega}{\partial \xi^2}&-& n_d^2 \frac{\partial^2
\Omega}{\partial \tau^2}+2 i \frac{\partial \Omega}{\partial \xi}+2
i n_d^2 \frac{\partial \Omega}{\partial
\tau} + (n_d^2-1) \Omega \nonumber \\
&&=3 \epsilon l \left(\frac{\partial^2 \rho}{\partial \tau^2}-2 i
\frac{\partial \rho}{\partial \tau}-\rho\right), \label{Maxdl}
\end{eqnarray}
where $\tau=\omega t$ and $\xi=kz$ are the dimensionless time and
distance, $\mu$ is the dipole moment of the quantum transition,
$\hbar$ is the reduced Planck constant, $\delta=\Delta
\omega/\omega=(\omega_0-\omega)/\omega$ is the normalized frequency
detuning, $\omega$ is the carrier frequency, $\omega_0$ is the
frequency of the quantum transition, $\gamma_{1}=1/(\omega T_{1})$
and $\gamma_{2}=1/(\omega T_{2})$ are the normalized relaxation
rates of population and polarization respectively, and $T_1$ ($T_2$)
is the longitudinal (transverse) relaxation time. The dimensionless
parameter $\epsilon= \omega_L / \omega = 4 \pi \mu^2 C/3 \hbar
\omega$ is responsible for the light-matter coupling, where $C$ is
the density of two-level particles and $\omega_L$ is the normalized
Lorentz frequency. Quantity $l=(n_d^2+2)/3$ is the local-field
enhancement factor due to the polarization of the background
dielectric with refractive index $n_d$ by the embedded quantum
particles \cite{Crenshaw}. Further we numerically solve Eqs.
(\ref{dPdtau})--(\ref{Maxdl}) using the finite-difference
approach adopted to this particular problem (see \cite{Novitsky2009, Novitsky2018}). Here we
limit the consideration to the one-dimensional problem described above as
it is usually done in analysis of self-induced transparency and similar effects.

Hereinafter we consider the 1D model of disorder \cite{Novitsky2018}: the density of two-level particles experiences
periodical random variations along the light propagation direction
as depicted in Fig. \ref{fig1}, so that the strength of light-matter
coupling in the $j$th period (layer) of the medium corresponding to
$(j-1) \delta L < z \leq j \delta L$ is given by
\begin{eqnarray}
\omega^{(j)}_L = \omega^0_L [1+2 r (\zeta_j-0.5)], \label{randvar}
\end{eqnarray}
where $\omega^0_L$ is the constant determined by the average density
of two-level particles, $\zeta_j$ is the random number uniformly
distributed in the range $[0; 1]$, and $r$ is the parameter of
the disorder. In other words, the system can be considered as a
multilayer (total thickness $L$) consisting of slabs of thickness
$\delta L$ with the different density of quantum particles. We should emphasize that we do not deal with the light scattering on individual randomly distributed particles. This could require the consideration of light-matter interaction and localization in 3D setting as, e.g., in Refs. \cite{Skipetrov2014,Skipetrov2018}. On the contrary, we analyze the light interaction with averaged ensembles that is a common practice for self-induced-transparency studies and, therefore, we can limit ourselves to more simple 1D problem. However, our model is a step forward comparing to the uniform-density resonant media usually considered in the literature (see, e.g., the classic works \cite{McCall,Poluektov,Allen}). We used similar model previously in \cite{Novitsky2018} to demonstrate the transition from the self-induced transparency to a localization
regime in the case of a single light pulse propagating through the disordered medium.

We should also note that we neglect the optical instability effects, including structural changes of materials and pattern formation under high-intensity illumination \cite{Rosanov2002,Arecchi1999}. The reason is that the optomechanical nature of such structural changes requires that the parts of the system can freely move as is the case for a cloud of cold two-level atoms \cite{Camara2015}. However, in our study, the quantum particles are assumed to be fixed at their positions in space as takes place in the solid-state structures, such as quantum dots embedded in glass or polymer. We also should stress the coherent nature of light-matter interaction needed to get self-induced transparency and $2 \pi$ soliton formation. This condition distinguishes our system from the usual random absorbers developed, e.g., for photovoltaics applications \cite{Battaglia2012,Hao2014,Siddique2017}.

The parameters used for calculations correspond to rare-earth atoms or semiconductor quantum dots as the two-level particles \cite{Palik,Diels}. In particular, we take the relaxation times
$T_1=1$ ns and $T_2=0.1$ ns (which are much longer than the light pulse
duration), the exact resonance ($\delta=0$), and the Lorentz frequency $\omega^0_L = 10^{12}$ s$^{-1}$. From the latter value one can estimate the required average concentration of quantum particles. For example, the quantum dots have rather large transition dipole moments (several tens of Debyes \cite{Eliseev2000, Borri2002}), so that one can take relatively low densities: if $\mu=30$ D, then we have $C^0 \approx 3 \cdot 10^{17}$ cm$^{-3}$. The mean Lorentz frequency and, hence, concentration can be made even lower for longer incident pulses \cite{Novitsky2018p}. The full thickness of the
medium is $L=1000 \lambda$, whereas the thickness of the constant density
layers is $\delta L = \lambda/4$. These values enable a
localization in the system \cite{Novitsky2018}. Incident light pulses
have the central wavelength $\lambda=0.8$ $\mu$m and the Gaussian
envelope $\Omega=\Omega_p \exp{(-t^2/2t_p^2)}$, where the pulse
duration is $t_p=50$ fs. For these short pulses we
can safely neglect the influence of near--dipole-dipole
interactions between the two-level emitters, since the condition $\omega_L t_p \ll 1$ holds \cite{Novitsky2010}. We also neglect the inhomogeneous broadening assuming $t_p \ll T^\ast_2$, where $1/T^\ast_2$ characterizes its contribution to the full line width. The peak
Rabi frequency $\Omega_p$ is in the units of
$\Omega_0=\lambda/\sqrt{2 \pi} c t_p$ corresponding to the pulse
area of $2 \pi$ usual for self-induced transparency
solitons. In the case of quantum dots discussed above, this amplitude corresponds to the intensity of about $1$ GW/cm$^2$ and can be made lower for longer pulses. As to the background dielectric, let us first consider particles in vacuum ($n_d=1$) for the sake of simplicity. Then this proof of principle will be extended to the silica glass background ($n_d \simeq 1.5$).

\section{Transmission modulation: Proof of principle}\label{vacuum}

\begin{figure}[t]
{\includegraphics[scale=0.9, clip=]{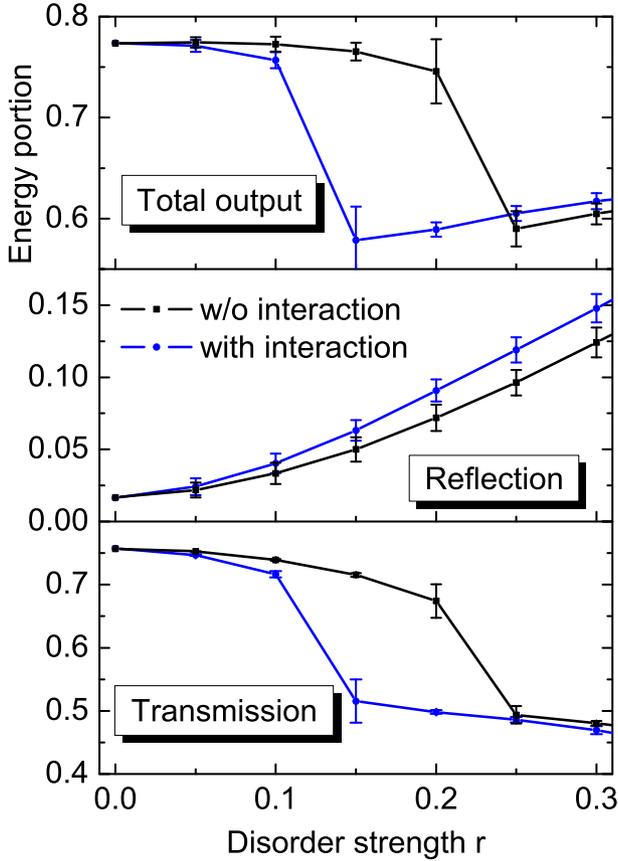}} \caption{\label{fig2}
Average transmitted, reflected and total energy of
pulses depending on the disorder parameter $r$. The cases of the
interacting co-propagating pulses and single pulses propagating without an interaction are shown. The amplitudes
of the pulses are $\Omega_{p1}=\Omega_0$ and $\Omega_{p2}=1.5
\Omega_0$; the interval between pulses is $\Delta t = 10 t_p$.
The energy averaged over $100$ realizations was calculated for the time
interval $700 t_p$ and was normalized on the input energy. The error
bars show the unbiased standard deviations for the corresponding
average values.}
\end{figure}

\begin{figure}[t]
{\includegraphics[scale=0.9, clip=]{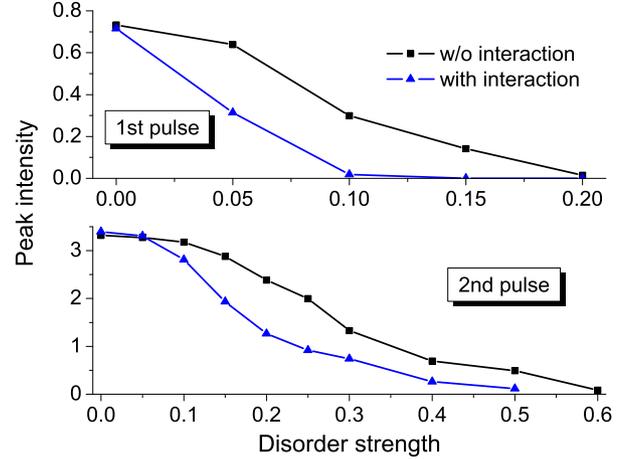}} \caption{\label{fig3}
Average peak intensities of the transmitted first and
second pulses depending on the disorder parameter $r$. The cases of the
interacting co-propagating pulses and single pulses propagating
without an interaction are shown. Peaks were averaged over $100$
realizations, other parameters are the same as in Fig. \ref{fig2}.}
\end{figure}

\begin{figure}[t]
\includegraphics[scale=0.75, clip=]{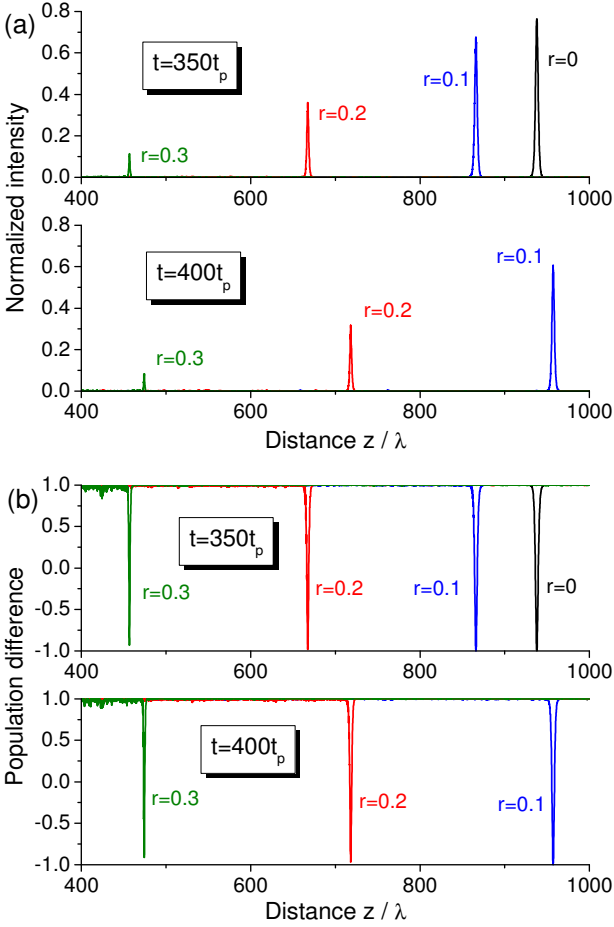}
\caption{\label{fig4} (a) Normalized intensity and (b) corresponding population difference at different moments of time shown for one disorder realizations at
different values of $r$. The parameters are the same as in Fig. \ref{fig2}.}
\end{figure}

\begin{figure}[t]
{\includegraphics[scale=0.9, clip=]{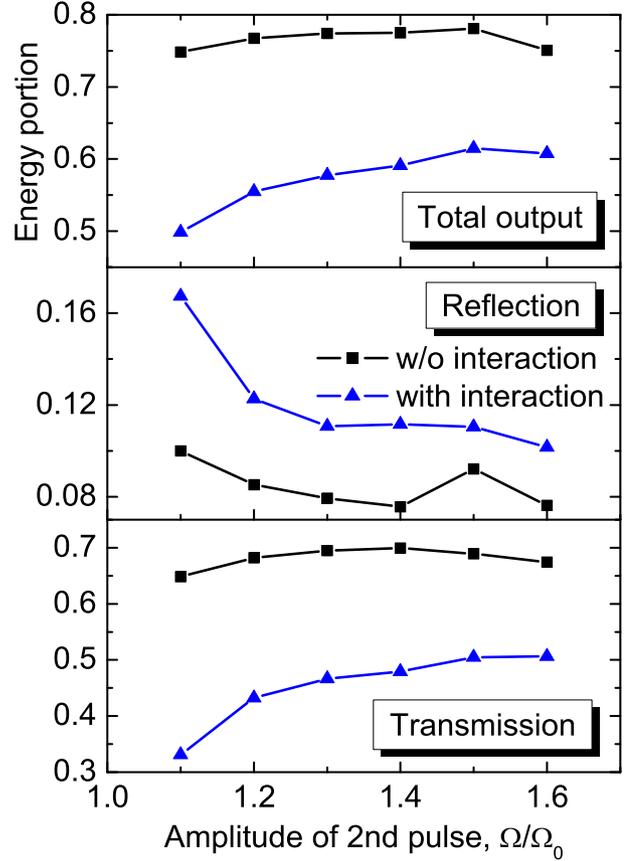}} \caption{\label{fig5}
The dependence of the transmitted, reflected and
total energy of pulses on the amplitude of the second pulse at the
disorder parameter $r=0.2$. The cases of the interacting co-propagating
pulses and single pulses propagating without an interaction are shown for one realization of the random medium, without averaging over a number of realizations as in Fig. \ref{fig2} and \ref{fig3}. The first pulse has the amplitude $\Omega_{p1}=\Omega_0$.}
\end{figure}

Let us consider the interaction of copropagating pulses having different
amplitudes (or areas). As mentioned above, in the ordered resonant
medium ($r=0$) such pulses eventually transform into standard $2
\pi$ solitons and interact elastically.
It is known that interaction of copropagating pulses can be realized due to the fact
that the pulse speed depends on its intensity: more powerful pulses move
faster \cite{Novitsky2011}. So, if the first pulse has lower amplitude than the second
one, they can collide and interact. This kind of interaction in disordered resonant media has not been considered previously to the best of our knowledge, being, however, of great importance for the modulation effect. In this section we consider two-level particles in vacuum ($n_d=1$), therefore, we can avoid a reflection at the interfaces, which would change the picture of the interaction otherwise. The practically realizable random structure with a dielectric host will be considered in the next section.

Let us consider Fig. \ref{fig2}, where the
dependencies of the transmission, reflection and their sum on the disorder
parameter $r$ are shown for the copropagating pulses with amplitudes
$\Omega_{p1}=\Omega_0$ and $\Omega_{p2}=1.5 \Omega_0$ and the inter-pulse
interval $\Delta t = 10t_p$ between peaks at the entrance. Here we
compare the case of two copropagating pulses with the
case of isolated pulses. The latter is used to calculate the
transmission and reflection for the non-interacting regime; namely, the transmitted or
reflected energy of two non-interacting pulses is
$E_{12}=(E_1 \Omega_{p1}^2+E_2
\Omega_{p2}^2)/(\Omega_{p1}^2+\Omega_{p2}^2)$, where $E_1$ and $E_2$
are taken from the isolated-pulse propagation calculations. It is
clearly seen from Fig. \ref{fig2} that the resulting transmission
and reflection for the pulses with and without interaction are the
same in the uniform medium ($r=0$) and are crucially different in the
presence of disorder ($r>0$). In other words, \textit{disorder
induces an effective interaction between copropagating pulses}, which
now collide inelastically losing some energy. The reason is
in the non-uniformity of the medium, therefore even slight
reflections or scattering on the density fluctuations of quantum particles increase the coupling between the pulses.

The additional loss due to the coupling between the pulses can be
used to control transmission of light through the system. Indeed,
after the collision, the pulses have lower intensity and, hence,
propagate slower becoming more vulnerable to trapping inside the
disordered medium. As a result, the transmission and the total output
demonstrate a drop at much lower disorder (see Fig. \ref{fig2}) associated with the shifting threshold of localization
\cite{Novitsky2018} to smaller values of $r$. Whereas for the interaction-free
regime this threshold occurs at $r \gtrsim 0.2$, for the
case of the interacting pulses the light is localized already at $r \gtrsim 0.1$. The situation considered can be treated as a peculiar scheme of \textit{transmission modulation}
based on the disorder-induced pulse-pulse interactions: launching the
second pulse significantly modifies transmission of light through
the medium, so that the transmission drops down to $0.5$ in comparison with the single-pulse transmission regime ($0.75$).

Let us consider in more details the evolution of the pulses inside the medium. Figure \ref{fig3} shows the change of the peak intensity of
the first and the second pulses transmitted through the medium with
different disorder parameters. Note that the second,
high-intensity pulse can pass through the medium for disorders $r \leq 0.5$, whereas the
first, low-intensity one disappears already at $r=0.1$ due to
interaction with the second one. The first pulse normalized intensity inside
the medium after the interaction with the second one is presented in
Fig. \ref{fig4}(a). The joint
action of the second pulse and disorder results in the strong decrease
of the first pulse intensity and, hence, its speed. The comparison
of the intensities and positions of the first pulse at 
$t=350 t_p$ and $400 t_p$ implies that it moves slower and loses
more energy with the increasing $r$. Finally, at $r=0.3$,
the pulse is almost motionless. At
the same time, this slow (almost standing) pulse is still very close
to the standard self-induced transparency soliton as shown in Fig.
\ref{fig4}(b). In particular, the population difference demonstrates the
usual pattern with the drop to the almost complete inversion and return back to the ground state, so that the energy absorbed on the leading front of the pulse is emitted by the medium on its trailing front \cite{McCall}.

Thus, the transmission modulation takes place because of the inelastic pulse-pulse interaction. As a result of this interaction, the pulses are losing additional energy and slowing down. This process leads to a stronger localization of light inside the medium.

\section{Optimizing parameters}\label{backgr}

\begin{figure}[t]
{\includegraphics[scale=1., clip=]{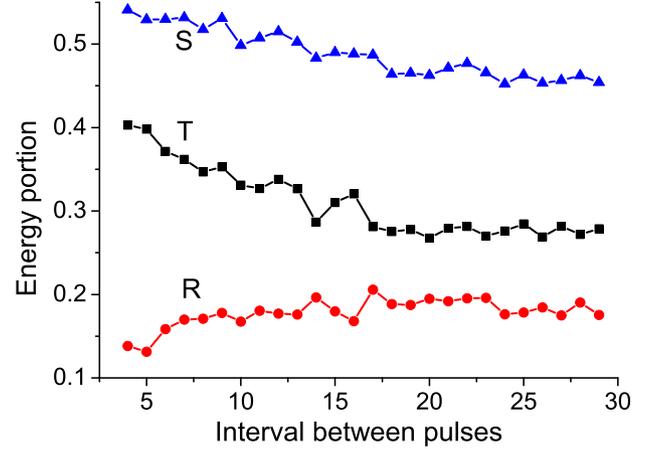}} \caption{\label{fig6}
The dependence of the transmitted, reflected and
total energy of pulses on the interval between pulses. The
disorder parameter is $r=0.2$, the amplitudes of the pulses are $\Omega_{p1}=\Omega_0$ and
$\Omega_{p2}=1.1\Omega_0$. The cases of the interacting co-propagating
pulses and single pulses propagating without an interaction are shown for one realization of the random medium, without averaging over a number of realizations as in Fig. \ref{fig2} and \ref{fig3}.}
\end{figure}

\begin{figure}[t]
{\includegraphics[scale=0.9, clip=]{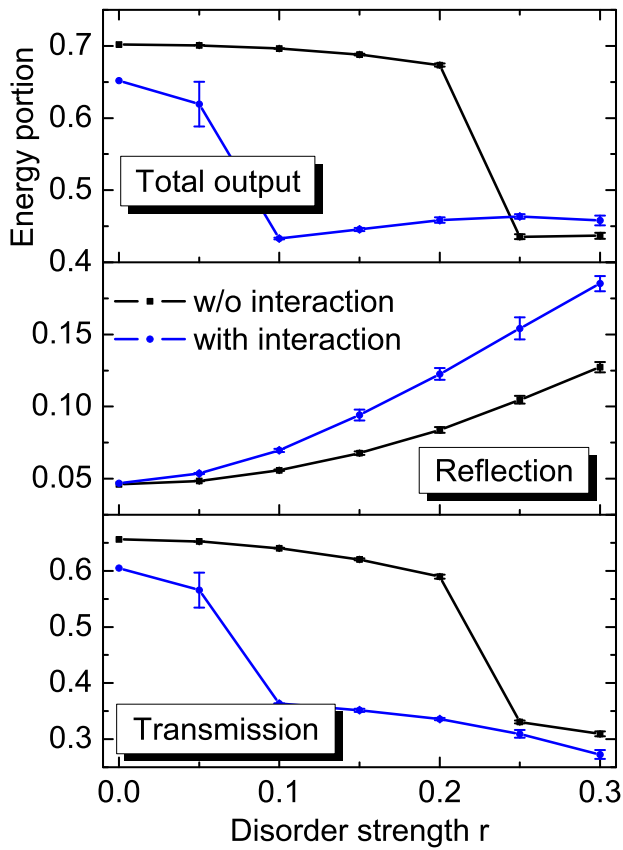}} \caption{\label{fig7}
Average transmitted, reflected and total energy of
pulses depending on the disorder parameter $r$ in the case of the
background refractive index $n_d=1.5$. The additional layers of pure background
material of thickness $40 \lambda$ have been added at the both sides of the medium to suppress the influence of reflection from the interfaces (see explanations in the text). The cases of the interacting co-propagating pulses and single pulses propagating without interaction are shown. The amplitudes of the pulses are $\Omega_{p1}=\Omega_0$ and $\Omega_{p2}=1.1 \Omega_0$. The energy averaged over $10$ realizations is calculated for
the time $700 t_p$ and is normalized on the input energy.
The error bars show the unbiased standard deviations for the
corresponding average values.}
\end{figure}

In this section, we analyze the influence of the pulse amplitude,
interval between pulses, and host dielectric on the transmission modulation discussed in the previous section. Let us
make the disorder strength equal to $r=0.2$ corresponding to the
strong influence of collisions on the transmission (see Fig. \ref{fig2}).
Figure \ref{fig5} shows the transmission and reflection for a single
realization of copropagating pulses without the aforementioned averaging. This is valid because of the small standard deviation at $r=0.2$, so that the transmission and reflection are approximately the same for every realization of disorder. The first pulse has the amplitude
$\Omega_{p1}=\Omega_0$. As it is expected, the interaction between the
pulses leads to a significant decrease of the transmission and a
rise of the reflection. Moreover, decreasing the amplitude of
the second pulse from $\Omega_{p2}=1.5 \Omega_0$ down to $1.1
\Omega_0$ leads to the transmission drop from $\sim 0.5$ to $\sim 0.3$, so
the ratio of the transmission of the pulses with and without
interaction (contrast factor) grows from approximately $1.5$ up to
$2$. Therefore, the pulses with close amplitudes
($\Omega_{p1}=\Omega_0$ and $\Omega_{p2}=1.1 \Omega_0$) are more
suitable for the transmission modulation, because they propagate with close velocities and, consequently, have more time to interact than colliding pulses with strongly differing velocities.

The second important parameter to be considered here is the interval
between pulses. Previously, we assumed it to be $\Delta t=10 t_p$.
The transmission and reflection dependencies on the interval are shown in Fig. \ref{fig6} (only one realization is demonstrated again). The calculations were performed for the pulses with the amplitudes $\Omega_{p1}=\Omega_0$ and $\Omega_{p2}=1.1 \Omega_0$ and
the disorder strength $r=0.2$. We see that for shorter intervals, the
transmission increases, i.e. they collide soon after launching
and lose less energy. On the contrary, increase in
$\Delta t$ first lowers transmission, and then, after $\Delta t \geq 20
t_p$, keeps practically the same value below $0.3$. Thus, the effect of
of the transmission modulation occurs in the wide range of intervals between pulses.

All the previous results are obtained for quantum particles in vacuum
($n_d=1$) and can be considered as a proof of principle. Hereinafter, to
extend the idea of the transmission modulation to a more
realistic scenario, we simulate the propagation of pulses in a glassy or polymeric medium ($n_d=1.5$) with the randomly changing density of quantum particles. The
crucial difference of this disordered structure from the aforementioned particles in vacuum is the additional reflections at the front and rear interfaces of the slab. These reflections can strongly influence the results even in the ordered case, since a part of the faster pulse reflected at the rear interface can interact with the slower
one. This situation resembles the case of counterpropagating pulses
interacting inelastically and leads to the trapping and absorption of a large amount
of energy in the medium (see also the next section). In order to
suppress this effect and preserve the copropagating geometry, we
place the additional layers of the pure host material (without two-level particles) of
thickness $40 \lambda$ to the both sides of the random medium. This is enough
to obtain almost interaction-free propagation of pulses at $r=0$, since the second pulse now collides with the reflected part of the first one outside the region with the resonant particles. The results of the transmission and reflection simulations averaged over $10$ realizations (enough to obtain reliable results) for the different disorders are shown in Fig. \ref{fig7}. The curves are generally similar to those shown in Fig. \ref{fig2} proving the possibility of the transmission modulation with the help of practically realizable, solid random structures.

\section{Note on counterpropagating pulses}\label{counter}

\begin{figure}[t]
{\includegraphics[scale=1., clip=]{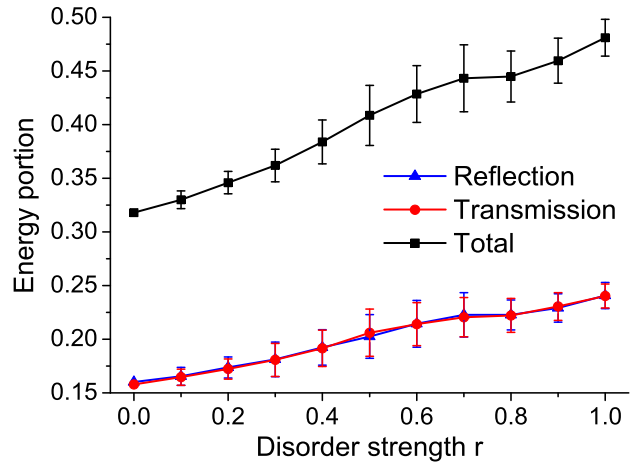}} \caption{\label{fig8}
(Color online) Average transmitted, reflected and total energy of the
interacting co-propagating pulses depending on the disorder
parameter $r$. The amplitudes of the pulses are
$\Omega_{p1}=\Omega_{p2}=1.5 \Omega_0$. Energy averaged over $100$
realizations is calculated for the time $500 t_p$ and is
normalized on the input energy. The error bars show the unbiased
standard deviations for the corresponding average values.}
\end{figure}

The previous sections are devoted to the copropagating pulses
interacting in a disordered resonant medium. The counterpropagating pulses, as we have already mentioned above, interact inelastically even in the ordered case \cite{Novitsky2011}. The introduction of the disorder to the structure does not change the situation significantly. The results of the calculations for colliding $3 \pi$ pulses are shown in Fig.
\ref{fig8} (we use the terms ``reflection'' and ``transmission'',
although the design is symmetric and they can be interchanged).
These results show that the disorder provides quantitative changes only, since significant portion of energy is trapped already at $r=0$ and the randomness can only slightly
increase the transmission and reflection.

\section{Conclusion}\label{concl}

In this paper, we analyzed in details the collisions of pulses in
a disordered resonant medium - random metamaterial - being a host transparent medium filled with two-level quantum particles, such as quantum dots or atoms. We have revealed and analyzed in details a new mechanism for an all-optical transmission modulation based on the inelastic interaction of copropagating pulses taking place due to the presence of the disorder in the system. The crucial parameters enabling the modulation tuning are revealed and considered. Apparently, a setup utilizing counterpropagating pulses doesn't provide this peculiar behavior due to the weak influence of the disorder on the interactions between such pulses. This novel approach for light control could be utilized for some highly demanded applications, such as modulation and switching of pulsed radiation.

\begin{acknowledgements}
The work was supported by the Belarusian Republican Foundation for Fundamental
Research (Project No. F18-049), the Russian Foundation
for Basic Research (Projects No. 18-02-00414, 18-52-00005, and
18-32-00160), Ministry of Education and Science of
the Russian Federation (GOSZADANIE, Grant No. 3.4982.2017/6.7), and
Government of Russian Federation (Grant No. 08-08). Numerical
simulations of light interaction with resonant media were supported
by the Russian Science Foundation (Project No. 17-72-10098). The dependencies of transmission on disorder parameters were calculated with partial support of the Russian Science Foundation (Grant No. 18-72-10127).
\end{acknowledgements}

\end{document}